\documentclass[fleqn]{article}

\newcommand{\tab}{\hspace{5mm}}

\begin{document}

\newcommand{\keywords}{photoelectric effect, Maxwell-Hertz electromagnetic theory,
detected signal} 
\newcommand{\PACS}{34.10.+x, 03.50.Kk, 84.47.+w, }
\newcommand{\email}{\tt vesely@itba.mi.cnr.it, \tt vesely@tana.it} 

\title{Concerning Hertz' photoelectric effect}
\author{S.\ L.\ Vesely$^{1}$, A.\ A.\ Vesely} 

\newcommand{\address}
  {$^{1}$I.T.B., C.N.R., via Fratelli Cervi 93, I-20090 Segrate(MI)
   \\ \hspace*{0.5mm} Italy \\ 
   }

\maketitle

{\small
\noindent \address
\par
\par
\noindent email: \email
}
\par

{\small
 \par
\par
\noindent{\bf PACS:} \PACS 
}

\begin{abstract}
\noindent Experimental evidence of the photoelectric effect 
goes back to H. Hertz. It occurred during the famous confirmation 
experiments of the Maxwellian theory. It is commonly held however 
that it cannot be explained in the framework of that theory. 
We are calling attention to some aspects linked with the interpretation 
of that effect on which, in our opinion, it is worthwhile reflecting.
\end{abstract}

\section*{Introduction}

As known, in 1905, in addition to the theory of relativity and 
a kinetic treatment of Brownian movement, A. Einstein proposed 
a heuristic explanation of the photoelectric effect \cite{Einstein1905}.
 We see in this proposal at least two original choices.

In the first place, in the 1905 work Einstein takes up a hypothesis 
of M. Planck, new and anything but accepted; it is he who will 
elucidate the innovative content thereof \cite{Einstein1909}. 
The hypothesis reproposed admits that upon thermodynamic equilibrium 
light be emitted and absorbed by matter only in discrete quantities, 
the so-called quanta. The innovative content of the hypothesis 
concerns energetically corpuscular constitution of light if we 
may express ourselves thus.

In the second place, he applies this hypothesis, and hence the 
theory in whose framework it is formulated, to a curious behaviour 
of metals and gives a simple explanation thereof. The experimental 
analysis of the effect on which he bases himself is P. Lenard's.

Today, the experiments of R. Millikan are taken as confirmation 
beyond a doubt of that theoretical hypothesis \cite{Stuewer2000}. 
It seems that, at the most, physics will try to improve the fits 
of foresight of the numerical values sanctioned therewith.

Here we would like to draw attention to the fact that no experiment 
verifying equations necessarily proves the correctness of a general 
theory. Admitted that a theory is sufficiently general and logically 
coherent, it suggests but does not guarantee a relationship between 
observed facts. Another theory, based on equally rigorous mathematics, 
could lend itself to connecting very similar facts in a manner 
at all different. We shall seek to show how this takes place 
in a specific case. That is to say we see how electromagnetism 
explains the photoelectric effect. We believe indeed that just 
this effect supplied H. Hertz with the key for testing Maxwell's 
theory experimentally.

\section*{Relationship of photoelectricity to electromagnetism}

The photoelectricity explained by Einstein pertains to the electrical 
behaviour of specifically irradiated metallic bodies. This is 
a physical fact, i.e. an observed fact, of which an objective 
description can be given. Under well defined experimental conditions 
it is reproducible.

Electromagnetism is a theory. Several theories have been proposed 
to account for electrical and magnetic phenomenology and Maxwell 
debates it in his treatise \cite{Maxwell1891}. The theory which we 
understand is his, i.e. the one summarized in the equations which 
bear his name. According to his intentions this theory accounts 
for the electrical facts observed by M. Faraday. Electromagnetism 
is notoriously also one of the theories of light phenomena.

Electromagnetism is a good theory if it is coherent and if it 
accounts for the physical facts for which it was written.

Verification of coherence is the duty of mathematical logic. \textit{As 
an alternative}, especially in physics, coherence has been proven 
by recourse to a model or construction of one. In this case the 
model places the observed facts in a certain relationship. In 
addition to substituting (partially) the logical reduction, it 
can act as a link for the physical explanation.

In this work we take for granted that Maxwell's linear equations 
allow development of a coherent general theory and we concern 
ourselves with the electromagnetic explanation of the photoelectric 
effect. Disregarding interpretative preferences, those who believe 
the electromagnetic theory insufficient or inadequate for explanation 
rule out this effect's being related to those allowed for by 
Maxwell or even repute that the incorporation of one more law 
formulated in such a manner as to characterize the phenomenon 
would make the theory incoherent.

Our point of view on the question being stated, we propose to 
trace the phenomenon most similar to the photoelectric effect 
which Maxwell surely considered again. For this purpose we state 
specifically that the paternity of the photoelectric effect is 
not attributed to P. Lenard but to H. Hertz.

Lenard \cite{Lenard1900} writes, ``the action of ultraviolet light 
[on metallic surfaces] was discovered by Hertz. Hallwachs was 
the first to show that there is a simple relationship [of UV 
lighting] with extraction of negative electricity [from the volume] 
of bodies. Today that relationship is well studied''. The experiment 
concerns current through a phototube when the cathode is irradiated.

Hertz writes \cite{Hertz1894}, ``Fast electrical movement excitation 
experiments being initiated I noticed an interaction [between 
emitting RLC and receiving RLC] which caused simultaneous release 
of sparks [between the facing capacitor spheres. \dots ] For some 
time I asked myself if I hadn't run into a new form of electricity 
capable of interacting at a distance. [\dots ] The phenomenon was 
later investigated by Righi, Hallwachs, Elster and Geitel; a 
satisfactory explanation of the mechanism is still lacking''. 
The experiment concerns metallic reflection in the UV.

Since by scientific convention the same effect is concerned
 \cite{Houstoun1938}, to say the least it is not characterized by detection 
with the electrometer of the convection current emitted. In Hertz' 
work indeed all mention of measurement of a current is lacking. 
So let us leave aside consideration of the mathematical expression 
for the electrical responsivity of the irradiated metallic surface 
and let us observe that photoelectricity is enumerated among 
electrical oscillation resonance phenomena. Those experiments 
were designed to prove or confute Maxwell's theory and they proved 
it.

It seems to us that the circumstance of confirmation presupposes 
some convergence of the observations of Hertz and those made 
previously by Faraday.

Faraday's electrical research was centered on transmission. Moreover, 
it is rich with insights into phenomenology, even spectacular, 
of metallic surfaces; His work also contains a crop of observations 
on the behaviour of dielectric materials called theretofore electrical 
insulators. When the author concerns himself with transmission 
through the air he distinguishes the spark release \textit{before} 
closing of the electrical contact in the mercury switch from 
the one released when the contact is broken. The second fact 
was called by him `self-induction'. The first \cite{Faraday1834}, 
described as a fast conduction trigger without metallic contact 
and without arc formation, is attributed by him to a tension 
state dependant on the chemical-physical conditions of the metallic 
surfaces involved. Both mean that the purple part of the band 
is transmitted between the metals in the `tension state' without 
the need of an antenna. Naturally Faraday knew nothing about 
antennas.

After Hertz, Lenard and especially Millikan, tended to exclude 
the contribution of reflection to the current, treating the metallic 
surfaces of the electrodes in such a manner as to make them nonspecular. 
They can, for example, be blackened and in this case it again 
depends on to what degree of blackening one wishes to refer the 
photoelectric effect on the underlying metal. Lenard speaks almost 
immediately of the photoelectric effect of retort carbon irradiated 
with the carbon arc \cite{Lenard1902}.

Although the application capacity varies considerably, the difference 
between pulse propagation in a dispersive medium and transmission 
of a modulated carrier in a range of frequencies such that the 
behaviour of the air approaches vacuum does not affect the possibility 
of electromagnetic representation. Thus it seems that Maxwell's 
theory should lend itself to some linear schematization of the 
photoelectric effect.

\section*{Modeling interpretation of the radiation-matter interaction}

In the previous paragraph we separated the problems of mathematical 
coherence from those linked to the plausibility of interpretation. 
We mentioned that in a physics theory the model serves to facilitate 
mathematical verification and we sought to make it plausible 
that with or without the photoelectric effect electromagnetism 
has to admit the same model.

But given the mathematics with which it is intended to face the 
problem the model is a more functional structure for explanation 
than is logical verification. One model can facilitate the interpretation 
of certain facts and another can better justify others.

The model adopted by Maxwell is hydrodynamic \cite{Maxwell1888}; 
for him the horizon of physics stops with matter and motion. 
Therefore, from his viewpoint, the model chosen does not prejudice 
understanding of the phenomena described. His successors were 
unable to endorse it because in the first place it forced him 
to disseminate his treatise with caveats against analogies with 
the behaviour of fluids, then because he let himself be led into 
a misunderstanding, believing he could attribute to field variables 
a \textit{direct dynamic content}.

After him, all attempts to include electrical and magnetic 
phenomena en bloc among those explainable by mechanics being 
exhausted, the problem of understanding the relationship between 
the field variables \textbf{E}\textmd{(x,y,z,t),} \textbf{D}\textmd{(x,y,z,t),}
 \textbf{B}\textmd{(x,y,z,t), 
and} \textbf{H}\textmd{(x,y,z,t)} and matter and therefore mechanics was posed.

This is a problem of interpretation. It cannot be solved by use 
of an ad hoc ether. It clearly shows the limit of Maxwell's model 
but does not give indications that the mathematical basis is 
insufficient. Thus H. Lorentz corrects the existing theory
 \cite{Lorentz1915} by postulating that the kinematic behaviour of electrically 
charged ponderable bodies gives the \textit{only measurable proof} 
of the presence of electromagnetic fields. He adjusts the kinematic 
behaviour of test bodies in the fields by adding to Maxwell's 
equations the dynamic law which bears his name. Hereinafter we 
consider his work with exclusive reference to the problem set 
here.

According to us, this author's main purpose was to credit the 
Maxwell equations no matter how by establishing conventions acceptable 
to the majority of the scientific community. Just because the 
burden of proof of the electrical fields is conceived according 
to gravitation it is not a matter of an explanation of a specific 
electrical experiment but of the interpretation of mathematical 
formulas. If it is remembered that retarded potentials are an 
excellent mathematical artifice it can be concluded that the 
model proposed as an alternative to hydrodynamics does not seek 
to be innovative.

If Lorentz set himself the above mentioned purposes the interpretative 
capacity of his model was distorted. Further analyzing other 
implications of Lorentz' model is beyond the scope at hand. 

In this paragraph we are not discussing the experimental evidence 
in favor of the electron but rather the appropriateness of modeling 
the radiation-matter interaction in electromagnetism with its 
help. From this point of view the electron hypothesis was judged 
insufficient at least by all the `founding fathers' of quantum 
mechanics to account for the experiments. Among these A. Sommerfeld 
observes that in classical wave theory the electrons are considered 
in continuous interaction with the ether and that every variation 
of their movement involves irradiation of waves. After which 
he says \cite{Sommerfeld1919}, ``One imagines an electron in the 
source point of every spherical wave which generates the electromagnetic 
field of the spherical wave according to the rhythm of its movement. 
[\dots ] Instead of speaking of an electron we should speak of 
a solution of Maxwell's equations which would express the conditions 
of atom-ether coupling in force upon irradiation''. Although 
he viewed the interpretation in the framework of a mathematical 
construction \textit{different} from Maxwell's, we believe the color 
change pointed out is of a more general interest. It means that 
to have sense physically the electromagnetic fields need not 
represent \textit{substance or energy}. They can represent a relationship 
between objects, for example an electrical coupling. The necessity 
of representing this relationship by means of a model arises 
because of the generality and abstractness of the theory.

In effect, the number of material manifestations which we can 
trace back to electrical phenomenology is not small. In addition 
to the movement of indices or movable galvanometer elements, 
to detect electrical phenomena we use sounds, photochemical and 
electrolytic reactions, and even calorimetry. Tracing back all 
these manifestations to a single type of abstract mechanical 
movement does not mean that we have accounted for the mechanism 
of the manifestations but rather that we intend to adopt a kinetic 
model for electromagnetism.

We do the contrary of that when we seek to relate all the above 
mentioned material manifestations to electromagnetism in a unitary 
manner. As everyone knows Faraday believed electricity to be 
connatural with the structure of matter. This hypothesis is faithfully 
reproduced in Maxwell's equations with the result of offering 
a theory with a very new conception and different from all the 
preceding mechanical ones.

Accepting interpretation of the fields as a relation and implementing 
Maxwell's equations with a \textit{non-mechanica}l model, electromagnetism 
appears as a physical theory explaining the body of Faraday's 
experiments without attributing any dynamic meaning to the variables. 
Mathematics describes the electrical and lighting relationship 
and the explanation of what we have to represent with the relationship 
depends on the model.

\section*{The object of electromagnetism is established by analogy}

In the previous paragraph we saw that the electromagnetic theory 
is difficult to interpret because Maxwell did not allow for anything 
but electricity and light. But the only experimental manner of 
acceding to these magnitudes is to transduce them. To clarify, 
we do not observe light with the eyes but its manifestation, 
i.e. luminous or illuminated objects.

As electromagnetism is not a theory of electrical transduction, 
it would readily be abandoned today if the interpretative problem 
had not had an experimental solution. The experimental solution 
is due to Hertz but H. Helmholtz had a primary role therein. 
The latter author, as usual for his time, felt that all electromagnetism, 
and not only the single effect, could be verified by analogy 
with an appropriate mechanism. He decided to draw a parallel 
with the world of sounds by comparing resemblances and differences 
at the \textit{propagation} level.

The acoustic analogy was heuristic in the sense that Einstein 
wished for his own interpretation of the photoelectric effect. 
Before Hertz carried out the first of the demonstrative experiments 
of the propagation of electrical force in the ether, his teacher, 
Helmholtz, had laid down for acoustics an experimental methodology 
which has defied time; to evaluate sound fields around acoustic 
sources he introduced test bodies which he called resonators. 
As we shall see in paragraph 6 below, Hertz transposed the investigation 
method exactly as it was from musical acoustics and adapted both 
field generators and test bodies to the technical requirements 
of electricity. Since not even Hertz knew anything about antennas, 
we cannot assert that we would judge his field probes according 
to present day parameters. We can only observe that Hertz' so-called 
dipoles are \textit{not} the experimental equivalent of the test body 
prescribed by Lorentz.

We warned that it is possible to doubt the functionality of Lorentz' 
electron as a field probe without detriment to the validity of 
Maxwell's equations. The experimental confirmation given by Hertz 
independently of the particular probe (provided it were suitable) 
used by him can be considered also.

In the paper, \"{u}ber die Beziehungen zwischen Licht und 
Elektrizit\"{a}t \cite{Hertz1889} Hertz explains what the experimental 
confirmation of electromagnetism consists of. ``[\dots  between 
light and electricity there is] a succession of delicate interactions 
like rotation of the polarization plane with the current, or 
variation of conduction resistance with light. In these [examples] 
however there is no direct transduction between light and electricity 
but the ponderable matter mediates between these two manifestations. 
We shall not concern ourselves with this group of phenomena [i.e. 
those of transduction]. There are stricter relationships between 
light and electricity. I shall defend this statement before you: 
light is an electrical phenomenon''.

We must not understand identification of light with electricity 
as an imposition of Hertz on nature. Rather it concerns the solution 
given by him to the problem of establishing what \textit{physical 
object} electromagnetism is concerned with. Today we might say 
that, with this identification, he inaugurated the radiotelephony 
era for the purposes of human telecommunications.

The possibility of obtaining and giving information without having 
to establish direct contact is given \textit{naturally} and is absolutely 
not limited to electrical phenomena or to human beings. On the 
contrary, the possibility of neglecting the peculiarities of 
the detector or probe for the purposes of electromagnetic telecommunications 
is postulated by Maxwell and proven experimentally by Hertz.

We mentioned that an effect very similar to Hertz' photoelectric 
effect belongs to the experimental basis of Maxwell's equations. 
Hertz himself, in describing the fact, considers it marginal 
in comparison with the verification of electromagnetism which 
he has in mind. And yet the photoelectric effect is the first 
fact he detected with the experimental arrangement suited to 
that verification. According to the theory, the photoelectric 
effect gives rise to a signal. Neither the technology for reception 
thereof nor its decoding nor attribution thereto of a meaning 
are prescribed in electromagnetism.

\section*{Inconsistency of Einstein's heuristic approach with Millikan's 
verification}

If we understand, the energetic interpretation given by Einstein 
to the photoelectric effect is heuristic in the sense that he 
asks if it would not be possible to formulate a general theory 
of transformation of light in \textit{something else} in place of 
the theory of Maxwellian conception. The theory in the making 
would be oriented towards the transduction process and in this 
case would distinguish the spark from the negative electricity 
extracted. At least in this respect the new theory diverges from 
the Hertzian basis of telecommunications. But it could be a theory 
of electrical phenomena for the same reason. Indeed, even Faraday's 
original experimenting is much richer with facts and leaves a 
broad margin for schematization of electrical phenomena beyond 
theories of light.

Since the theory conceived by Einstein imposes an energy 
balance of transduction it must postulate the principle of conservation 
of energy and therefore is bound to be a linear theory. In accordance 
therewith the modality hypothesized for transformation of light 
in something else must be theoretically represented by simple 
processes such as absorption and emission \cite{Einstein1916}. The 
tolerance with which the linearity requirement is transferred 
to the experiment is not prescribed.

Today a coherent general theory alternative to electromagnetism 
exits. We shall not touch here on the question intrinsic to its 
field of application.

On the other hand the experimental solution given by Millikan 
does \textit{not} represent for the new theory what the Hertz verification 
represents for the classical one. In effect, in the era of this 
author the theory of electrical circuits was universally accepted 
and the photodiode was considered a circuitry component with 
full rights. Einstein's linear equation being reproduced experimentally 
he himself observes that he can agree with any linear theory 
that manages to justify it.

As mentioned above, metallic surfaces can display the photoelectric 
effect. However Einstein does not specify any surface characteristic 
when he gets the following heuristic law.

\begin{eqnarray}
eV = Nh\nu - P
\end{eqnarray}

in the very broad context of light production and transformation 
energetics. In the formula, e is the charge per gram-equivalent 
weight of monovalent ions, V is the potential to which the surface 
of the irradiated solid body is brought, N is the number of real 
molecules thereof, P is the potential with which electricity 
is held therein and is termed work function, and $h\nu$ is the quantized 
light energy corresponding to incident radiation with frequency 
$\nu$.

It seems to us that in the Einstein formulation two misunderstandings 
are introduced.

1.\tab 
Faraday's electrolytic ions, for each reagent, met the specified 
chemical composition to the electrolytic purity grade. Today 
we can obtain them chemically more or less pure than this and 
verify their properties. The degree of purity of the electron 
on the other hand has always been associated with metallic conduction.

2.\tab If, as we are led to think, in 1905 Einstein was not predicting 
L.A.S.E.R., he attributes to natural light model characteristics 
taken entirely from Maxwell's theory. Today, on the contrary, 
one should decide either that L.A.S.E.R. is not a black radiator 
or to which black bodies Planck's theory applies.

If the experimenter does not cautiously appraise the challenge 
placed on measurement he will have the alternative of an interpretation 
in agreement with electromagnetism also or a mistaken one. Obviously 
to discuss transformation or transduction of light in something 
else it is necessary to attack experimentally the question upstream 
of the place where the photoelectric effect pertains to an electromagnetic 
radiation and to a metallic electric current. Since concomitantly 
with electromagnetism it is established that electric phenomena 
can be referred objectively (i.e. measured) the upstream place 
where the question should be faced is the one in which the electrical 
effects are separated from their material manifestations.

We found no evidence that Millikan had tried to distinguish the 
so-called work function P from the Volta effect. He merely subtracts 
the second value in volts from the first. But the justification 
of the validity of the Einstein equation \cite{Millikan1916} not 
only ignores transduction at the measurement level but even neglects 
to include in the energy balance concomitant chemical transformations 
which have nothing to do with the measurement procedure chosen. 
The most considerable of these transformations is the formation 
of a patina on the cathode but it is not the only one.

If he were to reduce the \textit{non electrical} side effects by lowering 
the radiant power and/or using less inflammable cathodes and/or 
improving the vacuum in the cathode ray tube, he would verify 
Einstein's heuristic equation under the hypothesis of Maxwell's 
theory (i.e. regardless of the reception modality).

But Millikan might have reached the conclusion that, under the 
measurement conditions in which none of the `excess' side effects 
were still perceptible, neither would he have been able to confirm 
Einstein's equation numerically beyond all doubt. Thus he undoubtedly 
verified it but in the framework of what general theory?

\section*{The photoelectric effect mechanism}

Up to this point we have sought to say that the Maxwell-Hertz 
theory explicitly disregards the characteristics of transduction 
and in addition that even the theory proposed by Einstein, although 
it intends to appraise the transduction energy balance, must 
disregard the specification of a particular mechanism.

We are now proposing a mechanism which accounts for the photoelectric 
effect and saying that it is the one presently suggested in electromagnetism.

We shall proceed as follows. First we shall review Helmholtz' 
mechanical analogy, then we shall explain which function has 
the photoelectric effect in Hertz' experiment. Lastly we shall 
transfer to the UV excitation the same receiving and transmission 
modality observed for Hertzian waves. This is possible under 
the Maxwellian radiation hypotheses.

\subsection*{Helmholtz' intellectual contribution to Hertz}

Many mechanical systems can be made to vibrate under appropriate 
conditions. Oscillations are determined of such an entity as 
to endanger the integrity of the structure only in a few.

For example, if the wheels of the vehicle are not balanced, at 
certain speeds the steering wheel vibrates in an annoying manner 
around its axis. On the contrary, if they are balanced this does 
not occur at normal speeds. In an analogous manner a ship cannot 
exceed a certain cruising speed without the planking beginning 
to tremble. In addition to the creaking, the sailing ship's structure, 
differently from the vehicle, emits in creaking a sound similar 
to a note. In navigation it is believed that the wave represents 
a sufficient stress and other singing effects are not sought. 
In the construction of musical instruments, on the contrary, 
the aim is to obtain some effect of this kind from the soundbox \textit{while 
the vibration is sustained}.

A hammer blow given to a shroud stretched between two steel girders, 
even if it is perceptible, does not correspond to a musical sound. 
In addition, without a soundbox there is no intensity of force 
capable of stressing the rope in such a manner as to give rise 
to a musical sound. So apparently the function of the soundbox 
is to give sonority to the movement of a plucked, rubbed or struck 
tight chord.

Going a bit further, we can attribute to this the property of 
amplifying the vibration of the chord, possibly with distortion. 
Now the amplification consists of pulling the mass of contained 
air and the immediately surrounding mass in some movement. We 
can imagine that the soundbox has the faculty of adapting the 
movement of the chord to that of the air.

This experimental base does not allow enunciation of a physical 
law putting the vibration frequency of the chord in correspondence 
with the perceived note.

In the meantime if the chord is represented 'with distributed 
mass' it is a simple pendulum only in a first approximation.

But let us hypothesize that the chord vibrations are stationary 
and the isochronism is controlled stroboscopically. Secondarily 
the soundbox necessary for perception of the sound does not develop 
pendular motion. If we except organ pipes it doesn't have even 
by far a one-dimension structure.

In the third place the air filling the space behaves in general 
as an even worse oscillator.

We might even hazard the hypothesis that the air does not at 
all enter into oscillation during the sound. In particular, it 
seems to us illusory to think that smoke or fog or a flame would 
help demonstrating the movement more instantly or faithfully 
than does the soundbox.

Instead it is certain that the air behaves as a channel of transmission 
in relation to sound and consequently the individual without 
hearing defects perceives sounds and noises. We can agree to 
call both of them `signals' as long as the movement of the air 
itself does not become perceptible, which is what takes place 
for example during explosions. The signals are transduced into 
sounds by the ear. Under loose conditions we can ignore transduction. 
But a theory independent of the modality of transduction is a 
signal theory. Only in this kind of theory is it unimportant 
to call signals `sounds' or `noises'. Obviously Helmholtz implicitly 
makes this assumption \cite{Helmholtz1877}.

Now let us summarize the work of Helmholtz in relation to the 
experiment of Hertz. He considers the siren as an instrument 
capable of sustaining a certain note as long as desired. He associates 
with the pitch of tones perceived in the air the linear speed 
of rotation of the siren disk holes which produce them. These 
speeds are constant in modulus. It is understood at this stage 
that the frequency of rotation which equals the pitch of the 
tone supplies an objective measurement after disregarding the 
attacking and releasing transients. But it is the mere instrumental 
recording of a fact which, if we are musicians, we are capable 
of evaluating independently and equally objectively by ear. In 
a subsequent stage, making use of the instrument, we can `educate' 
the ear to hearing without being musicians.

In effect, having chosen 440Hz for the reference frequency \textbf{a'} 
attributed to the tuning fork, Helmholtz is able to reduce the 
musical chords to simple frequency relationships. In this manner 
it is seen that the quality of sound of the soundboxes as obtained 
by spectral analysis in accordance with Fourier is not simple 
and that the consonance or dissonance of a chord concerns the 
musical tone more than the pitch of a single partial tone, commonly 
the prime tone.

Studying the phenomenon better it is seen that as loudness increases, 
non-linear effects can occur. The non-linearity consists of the 
fact that we hear tones which the instrument did not produce. 
These tones which Helmholtz calls combinatorial tones are sums 
and differences of the tones emitted provided they are perceptible.

The linear case of these effects is perceived as beats. During 
the beats, sounds of similar pitch emitted by \textit{different} instruments 
interfere temporally. Today beats can be used for tuning; the 
slower they are the more similar is the pitch of the sound of 
the \textit{two instruments compared}.

It should be said that -

1.\tab 
Helmholtz evaluates the \textit{spectral} sound distribution in the 
space \textit{around} the siren by measuring with tuned sensors invented 
by himself the intensity of several persistent simple tones. 
Therefore he seems to locate the sounds in the room.

2.\tab Even though he calls these sensors `resonators' they cannot 
be excited until they \textit{oscillate freely} when the external 
stress is broken. They cannot be thought of as sound sources. 
Therefore, in the same manner as the ear or other soundboxes, 
they amplify selectively by coupling with the air.

\subsection*{Acoustic analogy due to the Hertz photoelectric effect}

Hertz possessed a battery-powered RCL secondary electrical circuit 
which he used as a transmitter and also an RCL with receiver 
function. For the purpose of emphasizing the photoelectric effect 
he fed the receiver in series with the same battery as the transmitter. 
A mercury switch was part of the primary circuit.

Hertz didn't feed the transmitter persistently during the experiment 
nor did he use the charge transient but the discharge \textit{transient}. 
Upon opening the switch, a spark went off at the spark gap between 
the transmitter capacitor plates, two small polished, perfectly 
clean metal balls. Simultaneously the spark went off at the receiving 
RCL spark gap provided the latter was less than approximately 
3 meters away. The two spark gaps also had to be facing each 
other and any object placed between them could prevent the spark 
to the receiver from going off, provided it completely hid a 
band beyond extreme violet. Hertz determined the frequency of 
this band by dispersion through a quartz prism. He named this 
phenomenon \textit{the photoelectric effect}.

A good quality factor Q was an essential requisite on the receiver. 
If the transmitter also had a good Q, a mutual reinforcement 
of the spark could be noted.

Let us consider the interpretation. Hertz thought that his secondary 
circuits were equipollent with soundboxes. But the acoustic box 
serves to both selectively amplify and broadcast music. On the 
contrary, modern broadcasting makes use of antennas distinct 
from oscillating circuits.

If in addition to tuning resonators in frequency with each other 
Hertz had adapted both to the transmission channel, he could 
have avoided consuming the power necessary to produce sparks 
and would have received a less noisy signal. In reality he needed 
to receive and transmit sparks because he had no lower frequency 
reception techniques available.

From the point of view of reception the spark transduces an electric 
signal and represents it so to speak. The spark produced by an 
electrical circuit on the other hand is not analogous to a simple 
tone nor to a musical sound. It can be said that having neglected 
the last consideration was fatal to the Hertzian interpretation. 
We shall attempt to repropose it while allowing for it.

Hertz precisely `is not concerned' with adapting the impedance. 
This is equivalent of believing \textit{a priori} the inductive coupling 
to be weak or even believing it to be of the order of magnitude 
of the acoustic one.

If the assumption were true the Hertzian radiofrequency (r.f.) 
receiver would function as a wavemeter, recording amplitude maxima 
or, more correctly, intensity maxima of a continuous wave (CW) 
emitted at the resonance frequency. A wavemeter is Helmholtz' 
resonator. When it is placed in the trough of a stationary wave, 
i.e. in a zone where interference does not produce cancellation, 
it amplifies a persistent sound, possibly introducing distortions. 
By analogy, Hertz prefigures to himself receiving a sequence 
of sparks timed at radiofrequency as long as wanted. In truth, 
Hertz' transmitter does not broadcast a continuous wave but a 
free induction decay (FID) \cite{Skeldon2000}. Hertz nevertheless 
prefigures to himself that the r.f. FID of a transmitting resonator 
gives rise to a goodly number of discharge sparks when fed at 
the highest power. On the other hand he observes \textit{a single} 
spark from the spark gap of the transmitter and, at the same 
time, the reception signal (a spark). So he concludes that the 
electric emission is really oscillating at radiofrequency but 
even \textit{more damped} than that of a common diapason.

The assumption is reasonable and the consequences plausible. 
Its falsification depends on the existence in trade of klystrons. 
In comparison with klystrons, any mechanical vibration is damped, 
even that of the diapason, despite the fact that it is the only 
musical instrument able to transmit a FID and which transmits 
it \textit{on condition it be adapted correctly to the air}.

If Hertz had considered this point and had adapted the impedance 
in addition to tuning the receiver to the transmitter, he would 
have noticed that not only the transmitter but the receiver also 
radiates when excited. In addition, under coupling conditions 
phase relationships are determined.

The symmetry of behaviour between electrical transmitter and 
receiver can be such as to make them indistinguishable as regards 
function. On the contrary, only diapasons at the same nominal 
frequency can be acoustically coupled during transients and in 
this case the beat phenomenon is recorded.

By analogy Hertz would have received a voltage or current signal 
corresponding to the envelope of interference working at low 
power and low frequency (amplitude demodulation). Since this 
procedure does not give rise to a spark he would have had to 
develop the reception circuit further.

The need to detect sparks arises from the absence of this circuit. 
So Hertz adjusts the transmitting circuit so that the overvoltage 
at the spark gap would produce them. According to Faraday the 
excess current corresponds to the extra current of opening of 
the switch but the sudden trigger of the light discharge depends 
on disconnection of the mercury from the circuit. The analogy 
given to the diapason would be a hammer blow which produces a 
clang even in the disadapted diapason. According to Helmholtz, 
to this clang must be attributed an origin different from the 
note \textbf{a'} due to the fact that it is emitted isotropically and 
is audible. According to him these are frequency combination 
tones typical of the metallic structure.

The mechanism identified by Helmholtz explains the emission of 
the spark by an electrical circuit \textit{by analogy}. The resonance 
coupling in the UV on the other hand is explained \textit{the same 
way} electrical oscillations of coupled circuits are explained, 
when the behaviour of linear electrical circuits is the model 
of electromagnetism. This allows the statement that the photoelectric 
effect is an inductive coupling in the UV. It is very selective, 
hence strong and in addition the requisites of adaptation for 
transmission in vacuum are met. This interpretation would not 
be possible if Hertz had made use of a coherer or a photodiode 
for reception of the spark.

\section*{Conclusion}

It seems to us that Einstein's proposal to develop formal structures 
capable of giving a simple description of the basic phenomena 
hasn't lost its appeal. It even seems to us that the photoelectric 
effect could be fairly considered a typical and fundamental phenomenon. 
It should not be surprising that in its experimental characterization 
according to Hertz it appears basic for electromagnetism.

Its interpretation is not elementary because it is given in the 
framework of a model which assumes simple behaviour of electrical 
circuits. In this case, indeed, its observation is associated 
with the non-linear behaviour of circuits.

We have showed that the photoelectric effect can be interpreted 
with the help of the analogy with sounds, and that it can be 
explained in the framework of Maxwell theory.

We believe however that the explanation of the photoelectric 
effect might be simplified starting from a different model of 
Maxwell-Hertz electromagnetism.

\end{document}